\catcode`\@=11
{\count255=\time\divide\count255 by 60
\xdef\hourmin{\number\count255}
\multiply\count255 by-60\advance\count255 by\time
\xdef\hourmin{\hourmin:\ifnum\count255<10 0\fi\the\count255}}
\def\ps@draft{\let\@mkboth\@gobbletwo
    \def\@oddhead{}
    \def\@oddfoot{\hbox to 7 cm{\tiny \versionno
       \hfil}\hskip -7cm\hfil\rm\thepage \hfil {\tiny\draftdate}}
    \def\@evenhead{}\let\@evenfoot\@oddfoot}

\def\citen#1{\if@filesw \immediate\write \@auxout {\string\citation{#1}}\fi%
\@tempcntb\m@ne \let\@h@ld\relax \def\@citea{}%
\@for \@citeb:=#1\do {\@ifundefined {b@\@citeb}%
    {\@h@ld\@citea\@tempcntb\m@ne{\bf ?}%
    \@warning {Citation `\@citeb ' on page \thepage \space undefined}}%
    {\@tempcnta\@tempcntb \advance\@tempcnta\@ne
    \setbox\z@\hbox\bgroup\ifcat0\csname b@\@citeb \endcsname \relax
    \egroup \@tempcntb\number\csname b@\@citeb \endcsname \relax
    \else \egroup \@tempcntb\m@ne \fi \ifnum\@tempcnta=\@tempcntb
    \ifx\@h@ld\relax \edef \@h@ld{\@citea\csname b@\@citeb\endcsname}%
    \else \edef\@h@ld{\hbox{--}\penalty\@highpenalty
    \csname b@\@citeb\endcsname}\fi
    \else \@h@ld\@citea\csname b@\@citeb \endcsname \let\@h@ld\relax \fi}%
\def\@citea{,\penalty\@highpenalty\hskip.13em plus.13em minus.13em}}\@h@ld}
\def\@citex[#1]#2{\@cite{\citen{#2}}{#1}}%
\def\@cite#1#2{\leavevmode\unskip\ifnum\lastpenalty=\z@\penalty\@highpenalty\fi%
  \ [{\multiply\@highpenalty 3 #1%
  \if@tempswa,\penalty\@highpenalty\ #2\fi}]}   %
\makeatother 

\def\alg           {algebra}
\def\Alpha         {N}
\def\auto          {automorphism}
\def\bA            {|{\cal B}^A\rangle}
\def\Bar           {\tilde}
\def\bc            {boundary condition}
\def\be            {\begin{equation}}
\def\bearl         {\begin{array}{l}}
\def\bearll        {\begin{array}{ll}}
\def\bfe           {{\bf1}}

\def\boundlb       {\langle {\rm B}_\lambda|}
\def\boundlk       {|{\rm B}_\lambda\rangle}
\def\boundok       {|{\rm B}_{\lambda=0}\rangle}

\def\calf          {{\cal F}}

\def\caln          {{\cal N}}

\def\cb            {chiral block}

\def\cft           {conformal field theory}
\def\cfts          {conformal field theories}
\def\chii          {\raisebox{.15em}{$\chi$}}
\def\chil          {\chii_\lambda^{}}
\def\chilc         {\raisebox{.15em}{$\breve\chi$}_\lambda^{}}
\def\chir          {\mbox{$\liefont W$}}

\def\corfu         {correlation function}
\def\Corfu         {correlation function}

\def\ctype         {Chan\hy Paton type}
\def\cvo           {chiral vertex operator}

\def\dl            {\mathbb }

\def\ee            {\end{equation}}
\def\eE            {{\rm e}}
\def\eear          {\end{array}}

\def\eq            {\,{=}\,}
\newcommand\erf[1] {(\ref{#1})}
\newcommand\exT[1] {#1}

\def\furu          {fusion rule}

\def\hil           {{\cal H}}
\def\hl            {\mbox{${\cal H}_\lambda$}}
\newcommand\hsp[1] {\mbox{\hspace{#1 em}}}

\def\hy            {$\mbox{-\hspace{-.66 mm}-}$}

\def\ii            {{\rm i}}

\def\iN            {\,{\in}\,}
\def\irmod         {irreducible module}

\def\jf            {J.\ Fuchs}
\long\def\labl#1   {\label{#1}\ee}
\def\Ldots         {,...\,,}

\def\lie           {Lie algebra}

\def\liefont       {\mathfrak }
\def\llb           {\mbox{\large(}}
\def\lrb           {\mbox{\large)}}

\newcommand\N[3]   {\mbox{$\caln_{\!#1#2}^{\;\ #3}$}}

\newcommand\nxt[1] {\\[.1em] \raisebox{.12em}{\rule{.35em}{.35em}}\hsp{.6}#1}
\def\om            {\omega}

\def\onedim        {one-dimen\-sional}

\def\ot            {\raisebox{.07em}{$\scriptstyle\otimes$}}
\def\oT            {\,\ot\,}
\def\otimeS        {\,{\otimes}\,}

\def\parfu         {partition function}
\def\pe            {\mbox{${\dl P}^1$}}
\newcommand\pHB[3] {\Psi^{#1#2}_{#3}}

\newcommand\pho[1] {\phi_{#1,\Bar{#1}}}

\newcommand\rb[3]  {R^{#1}_{#2 \Bar{#2};#3}} 
\newcommand\rc[3]  {R^{#1}_{#2 \Bar{#2};#3}}

\def\reals         {{\dl R}}
\def\rep           {rep\-re\-sen\-ta\-ti\-on}

\def\resp          {respectively}

\def\scs           {\scriptstyle}
\newcommand\sct[1] {\mbox{$ $}\\[-.5em]{\bf #1.}\\[.22em]}

\def\sss           {\scriptscriptstyle}
\def\tc            {\mbox{$\hat C$}}
\def\ttype         {automorphism type}
\def\twodim        {two-di\-men\-si\-o\-nal}

\def\vac           {0}

\newcommand\version[1] {\ifnum\draftcontrol=1 \typeout{}\typeout{#1}\typeout{}
                   \vskip3mm \centerline{\fbox{{\tt DRAFT -- #1 -- }
                   {\small\draftdate}}}
                   \vskip3mm \fi}

\def\wrt           {with respect to }

\def\wzwm          {WZW model}

\def\draftdate{\number\month/\number\day/\number\year\ \ \ \hourmin }

\global\def\draftcontrol{0}
\catcode`\@=12
\documentclass[12pt]{article} \usepackage{amssymb,amsfonts}
\begin{document}

\begin{flushright}  {~} \\[-15 mm]  {\sf hep-th/9801190} \\[1mm]
{\sf CERN-TH/98-17} \\[1 mm]
{\sf January 1998} \end{flushright}
 
\begin{center} \vskip 15mm
{\Large\bf D-BRANE CONFORMAL FIELD THEORY}\\[16mm]
{\large J\"urgen Fuchs} \\[3mm]
Max-Planck-Institut f\"ur Mathematik\\[.6mm] 
Gottfried-Claren-Str.\ 26, \  D -- 53225~~Bonn\\[11mm]
{\large Christoph Schweigert} \\[3mm] CERN \\[.6mm] CH -- 1211~~Gen\`eve 23
\end{center}
\vskip 20mm
 
\begin{quote}{\bf Abstract}\\[1mm]
We outline the structure of boundary conditions in conformal field theory.
A boundary condition is specified by a consistent collection of reflection
coefficients for bulk fields on the disk together with a choice of an
automorphism $\omega$ of the fusion rules that preserves conformal weights.
Non-trivial automorphisms $\omega$ correspond to D-brane configurations for
arbitrary conformal field theories.
\end{quote}
\vfill {}
\begin{flushleft}  {~} \\[-3 mm] {\sf CERN-TH/98-17} \\[1 mm]
{\sf January 1998} \\[3mm]
\noindent ------------------\\[1 mm]
{\footnotesize Slightly extended version of a talk given by 
J.\ Fuchs at the XXXI\,st International Symposium Ahrenshoop on the Theory of 
Elementary Particles (Buckow, Germany, September 1997)}
\end{flushleft}


\sct{String theory and \cft}
A complete understanding of string theory certainly requires many more
ingredients than just \cft, e.g.\ when it comes to finding a guiding
principle that would tell what solitonic sectors (and with which multiplicities)
must be included to arrive at a consistent theory. On the other hand, 
both at a conceptual and at a computational level, \cft\ does lead 
very far indeed. While at the level of string perturbation theory this is 
more or less accepted knowledge in the case of closed strings, it is a
prevailing prejudice that some of the more recently discovered structures that 
are tied to the presence of open strings with non-trivial boundary conditions
are inaccessible to \cft. This is of course a logical possibility, but
before making a decision on this issue one should better inspect the
tools that are summarized under the name `\cft' with sufficient care. 
I\exT{n the course of these investigations i}t may well turn out that 
present day knowledge about these matters is as yet incomplete and that the
uses of \cft\ can be largely expanded by further efforts.

Indeed we claim that the basic new features of open as compared to closed 
strings, such as e.g.\ D-branes (possibly with field strength, or multiply
wrapped) are well accessible to \cft. Moreover, once a suitable framework for 
\cft\ on closed orientable \exT{Riemann }surfaces ({\em closed\/} \cft, for 
short) is formulated \cite{fuSc6}, establishing the theory also 
on the open and\,/\,or non-orientable surfaces
({\em open\/} \cft) that arise as world sheets of open strings does not
pose any major conceptual problems any more, though there are several new
ingredients which considerably complicate matters at a more technical level.

\sct{Building blocks}
Let us first recall a few facts about the world sheet picture of closed
strings. The guiding principle for the construction of a string theory 
is to start with some given \cft\ (\exT{supposed to be }consistently defined 
on all closed orientable Riemann surfaces $C$) and then to discard the 
dependence on the properties of the world sheet $C$
while still keeping information about the field theory\exT{ on $C$}.
This is achieved by eliminating first the (super-)\,Virasoro algebra
via the relevant semi-infinite cohomology, then the choice of a 
conformal structure on $C$ via integration over the moduli space of complex
structures, and finally the choice of topology of $C$ by a summation over 
topologies. The latter sum is weighted by the power $\gamma^{-\chi}_{}$ of the 
string coupling constant $\gamma$, with $\chi\eq2{-}2g$ the Euler number of $C$.
In particular, string scattering amplitudes are obtained from
the $n$-point \corfu s $\calf_{\!g,n}\,{\equiv}\,\calf_{\!g,n}(\vec\lambda;
\vec z,\vec\tau)$ of the \exT{conformal }field theory by integrating over 
the moduli $\vec\tau$ of the genus-$g$
surface $C$ and (modulo M\"obius transformations) over the insertion points
$\vec z\,{\equiv}\,(z_1,z_2\Ldots z_n)$, and afterwards multiplying with
$\gamma^{-\chi}_{}$ and summing over $\chi$.

For a \cft\ to be consistently defined on all surfaces $C$, the \Corfu s 
$\calf_{\!g,n}$ have to satisfy various locality and factorization 
constraints. The former require that the $\calf_{\!g,n}$ are ordinary functions 
of the insertion points $\vec z$ and (up to the Weyl anomaly) of the moduli 
$\vec\tau$, while the latter implement compatibility with singular limits in 
the moduli spaces. These constraints are formulated in terms of the \cft\ on $C$
(which is orient{\em able\/}, but does {\em not\/} come naturally
as an orient{\em ed\/} surface), to which we refer as the stage of
{\em full\/} \cft. This stage must be carefully distinguished from the
stage of {\em chiral\/} \cft, where in place of the \corfu s one is dealing
with \cb s. Usually this stage is introduced by a somewhat heuristic
recipe for `splitting the theory into \exT{two }chiral halves'. A more
appropriate, and for the present purposes more convenient, description of
the chiral theory is as a \cft\ on an orient{\em ed\/} covering surface \tc\ 
which has the structure of a complex curve and from
which the \exT{original }surface $C$ can be recovered by dividing out an
anti-conformal involution \cite{fuSc6}.

For large classes of \cfts, in particular for \wzwm s, all correlation 
functions $\calf_{\!g,n}$ can in principle be computed exactly (i.e.,
fully non-perturbatively in terms of the field theory on the world sheet).
Moreover, in many interesting cases -- including, but by no means exhausted by,
free field theories -- at least at string tree level this can also be achieved 
in actual practice. The reason is that the \cb s can be obtained as the
solutions to the Ward identities of the theory. 
Let us note that even though \cft\ is typically formulated in an operator
picture, for establishing the Ward identities (and also for many 
other purposes) the existence of an operator formalism is not needed. 
Namely, the Ward identities
constitute identities for \cb s that can be formulated solely
in terms of the \rep\ theory of the relevant chiral \alg\ \chir, without
making use of an operator formalism. Also, once the \cb s are known, the 
\Corfu s are determined by the locality and factorization constraints, also 
known as sewing constraints, which (are believed to) possess a unique solution.
Of course, in string theory one usually interprets the scattering
amplitudes as expectation values for products of suitable vertex operators
for the string modes.
In \cft\ terms this amounts to working with an operator formalism, in which the
string modes are realized as (Virasoro-primary) \cvo s in the chiral, \resp\ 
as corresponding fields in the full theory. The locality and factorization 
properties constitute a necessary prerequisite for the existence of operator 
product expansions of the full theory. 

Via factorization, one can reduce many issues of interest to
statements about only a small number of building blocks, namely
the chiral 3-point blocks on \pe, and these building blocks can be studied
in terms of the \rep\ theory of the chiral \alg\ \chir. For instance, the 
index set $\{\lambda\}$ (an $n$-tuple of which labels the 
\exT{\corfu s }$\calf_{\!g,n}$\exT{, and which in the operator picture 
indicate the allowed fields}) corresponds to a suitable set $\{\hl\}$
of \irmod s of \exT{the algebra }\chir, and in rational theories the numbers 
$\N{\lambda_1}{\lambda_2}{\lambda_3}$ of independent 3-point blocks of type 
$(\lambda_1,\lambda_2,\lambda_3)$ are related, via the Verlinde formula,
to the modular behavior of the characters $\chil$ of these modules \hl.

For {\em open\/} strings, including {\em D-branes\/}, 
the situation is more complicated technically, but not conceptually. 
Some of the concepts \exT{mentioned above }are now realized 
in a somewhat different manner, but still they can be applied in much the 
same way as before. For instance, we have:
\nxt 
The Euler characteristic $\chi$ still counts the order in the string 
perturbation theory. But now $\chi$ is given by $\chi\eq2{-}2g{-}b{-}c$, where
$g$, $b$ and $c$ are the numbers of handles, boundary components, and
crosscaps of the surface $C$, respectively.
\nxt One must still distinguish between the two stages of the chiral and
the full conformal field theory. The full theory on $C$ can again be expressed 
in terms of  a chiral theory on some surface \tc\ by imposing locality and 
factorization constraints.
\nxt Again \tc\ is an oriented cover of $C$ from which one recovers
$C$ by modding out an anti-conformal involution $I$. But
now \tc\ is connected, whereas in the closed case it consists of two connected 
components each of which is isomorphic to $C$ as a real manifold \cite{fuSc6}.
Also, the involution
$I$ may now possess fixed points, giving rise to boundaries of $C$.
\nxt Again factorization allows to formulate the theory in terms of a few
building blocks. But besides the 3-point blocks on \pe, one now also needs
the 1-point blocks on the disk $D\eq\pe\!{/}\!_{z\mapsto1/z^*}$ as well as the 
1-point blocks on the crosscap ${\dl P}\reals^2\eq\pe\!{/}\!_{z\mapsto-1/z^*}$.

\sct{Boundary states and boundary conditions}
In contrast to the closed case, in open \cft\ \cite{card9,lewe3,prss3,fuSc6}
the locality and factorization constraints typically admit more than one 
solution, e.g.\ the 1-point \Corfu s $\langle\pho\lambda\rangle_{\!A}^{}$ of 
bulk fields $\pho\lambda$ on the disk $D$ depend on some additional label $A$.
These correlators are simply proportional to the corresponding 1-point blocks;
the constant of proportionality is the product of two factors
$\Alpha^{AA}_\vac$ and $\rc A\lambda\vac$. 
The number $\Alpha^{AA}_\vac$ is interpreted as the expectation value 
$\langle\pHB AA\vac\rangle$ of a `boundary vacuum field' \cite{card9}
$\pHB AA\vac$; roughly, the role of the boundary field is to make a geometric
boundary component into a `field theoretic boundary' that carries the 
boundary label $A$. Similarly, $\rc A\lambda\vac$ is a {\em reflection 
coefficient\/}, defined via the expansion \cite{lewe3,prss3}
  \be  \pho\lambda(r\eE^{\ii\sigma}) \,\sim\, \sum_\mu\sum_a
  (r^2{-}1)^{-\Delta_\lambda^{}-\Delta_{\Bar\lambda}+\Delta_\mu^{}}_{}\,
  \rc A\lambda\mu\,
  \pHB AA\mu(\eE^{\ii\sigma}) \qquad {\rm for}\;\ r\to 1  \labl{pp}
of $\pho\lambda$ in terms of boundary fields.  Every consistent collection of 
1-point correlators for all bulk fields, or equivalently, every consistent 
collection of reflection coefficients $\rc A\lambda\vac$, is referred to 
\cite{card9} as a {\em \bc\/} $A$. For free fields these amount to \bc s
in the ordinary geometric sense, but in the general case such an interpretation
is not available. Roughly, one can interpret the relation \erf{pp} by
imagining that to every bulk field there
is associated a kind of mirror charge on $\pe{\setminus}D$, which in turn
corresponds to some charge distribution on the boundary.

In the literature it is common to denote the 1-point chiral blocks on the disk 
by $\boundlk$ and to refer to them, as well as to their linear combinations 
  \be  \bA:= \sum\raisebox{-.45em}{$\scs\!\!\lambda$}\, \Alpha^{AA}_\vac\,
  \rb A\lambda\vac \, \boundlk \,,  \labl{bA}
as {\em boundary states\/}. Such an object is, however, not a state in the 
usual sense.  While formally it satisfies relations of the form
  \be  \llb W_n \oT\bfe - (-1)^{\Delta(W)}_{}\,\bfe\oT W_{\!-n} \lrb
  \boundlk = 0 \,,  \labl w
and in concrete examples can be written\,%
\footnote{~The formul\ae\ in the literature actually describe the
specific situation that the insertion point is at $z\eq0$ and that standard 
local coordinates on the covering surface $\tc\eq\pe$ of the disk are chosen.}
as an (infinite) sum of basis elements of the tensor product space
$\hil_\lambda^{}\otimeS\hil_{\Bar\lambda}$ of the relevant \chir-modules, 
it is {\em not\/} an element of that space, nor even of the completion of 
the tensor product space \wrt its standard scalar product.
Rather, the correct interpretation is indeed as a 1-point block on the disk.
At a more technical level, this can be described as a so-called 
{\em co-invariant\/} of the space $\hil_\lambda^{}\otimeS\hil_{\Bar\lambda}$
\wrt the action $W_n \oT\bfe - (-1)^{\Delta(W)}_{}\,\bfe\oT W_{\!-n}$
of the chiral \alg\ \cite{fuSc6}. 
In place of these somewhat unfamiliar objects one may equivalently consider
the singlets in the dual space 
$(\hil_\lambda^{}{\otimes}\hil_{\Bar\lambda})^\star_{}$;
thus roughly, the boundary states may also be regarded as genuine vectors in
this dual space. \exT{(Briefly, the notion of a co-invariant generalizes the
concept of a singlet-submodule to the case of non-fully reducible modules,
and the co-invariants of a module $\hil$ form a vector space isomorphic to
the singlets in $\hil^\star$.)}

In string theory, one often regards the boundary state $\bA$ as a synonym 
for the \bc\ $A$; its proper interpretation is that by
saturating one leg of a multi-reggeon vertex with $\bA$ 
amounts to introducing a boundary of type $A$ on the world sheet.
The quantities $\bA$ also appear naturally in the vacuum amplitude for the
annulus, which can be evaluated with the help of the formula
  \be  \boundlb \eE^{2\pi\ii\tau(L_0+\tilde L_0 - c/12)}_{} \boundlk =
  \chil(2\tau) \,,  \labl c
where $\chil(\tau)\,{\equiv}\,\chil(\tau{,}0{,}0)$ is the Virasoro-specialized
character of the \chir-module \hl\
(normalized, for convenience, to the quantum dimensions).

\sct{Twisted actions of the chiral \alg}
A basic task in open \cft\ is to determine all possible \bc s.
The properties to be imposed depend on the application \exT{that }one has 
in mind. In the context of \exT{\twodim\ }critical phenomena typically 
the boundary condition need to preserve just the Virasoro
algebra; in special situations it may even be sufficient to respect only 
part of it. In string theory, one commonly requires to preserve
the symmetry that is gauged, i.e.\ the Virasoro algebra
\resp\ its relevant super extension\exT{ in the case of superstrings}; 
but \bc s for which the (super-)Virasoro algebra is preserved only up 
to BRST-exact terms seem to be perfectly admissible as well.
Boundary conditions that violate part of the bulk symmetries can be roughly
imagined as describing boundaries that carry some charge already in the
absence of any fields.

The boundary blocks $\boundlk$ introduced above do preserve the full 
chiral \alg\ \chir. Here the precise sense of the term
`preservation' is that \chir\ acts
on $\hil_\lambda^{}{\otimes}\hil_{\Bar\lambda}$ as \exT{prescribed }in 
\exT{the formula }\erf w, 
i.e.\ the action on \exT{the second factor }$\hil_{\Bar\lambda}$ 
is twisted by the \auto\
  \be  \sigma_0:\quad W_n \;\mapsto\; (-1)^{\Delta(W)+1}_{}\, W_{\!-n}  \ee
\exT{ of \chir}. 
It is then natural to look for other chiral blocks that constitute co-invariants
for some differently twisted action of \chir. One way to achieve this is to
replace $\sigma_0$ by \exT{the product }$\sigma\,{\circ}\,\sigma_0$,
with $\sigma$ some other \auto\ of \chir.
One can check that (formal) solutions to 
  \be  \llb W_n \oT\bfe - (-1)^{\Delta(W)}_{}\,\bfe\oT \sigma(W_{\!-n}) \lrb
  \boundlk_{\!(\sigma)}^{} = 0   \ee
(which replaces the condition \erf w) are given by
$\boundlk_{\!(\sigma)}^{}\eq(\bfe \oT \theta_\sigma)\boundlk$,
where the map $\theta_\sigma$ which acts on $\hil_{\Bar\lambda}$ is 
characterized by its `$\sigma$-twining' property 
$\theta_\sigma{\circ}W_n=\sigma(W_n)\,{\circ}\, \theta_\sigma$.

Note that for non-trivial $\sigma$, such \bc s typically do {\em not\/} 
preserve the Virasoro \alg, and accordingly they shouldn't play a role in 
applications to strings.

As a side remark, we mention that a large class of \exT{examples for }Virasoro
non-preserving \auto s $\sigma$,
for which \exT{the induced map }$\theta_\sigma$ still has reasonable 
properties, is provided
by the \auto s $\sigma\eq\sigma_{\rm J}$ that implement \cite{fusS3} the 
action of simple currents J of \wzwm s. When such an \auto\
$\sigma_{\rm J}$ has order 
two, then e.g.\ analogues of the formula \erf c are given by
  \be  \bearl
  {}^{}_{(\sigma)\!}\boundlb \eE^{2\pi\ii\tau(L_0+\tilde L_0 - c/12)}_{} 
  \boundlk^{}_{\!(\sigma)} = \chil(2\tau{,}{-}\bar\varpi_{\rm J}\tau{,}
  (\bar\varpi_{\rm J}{,}\bar\varpi_{\rm J})\tau/2) \,,
  \hsp{2.2}  \\{}\\[-.55em]
  {}^{}_{(\sigma)\!}\boundlb \eE^{2\pi\ii\tau(L_0+\tilde L_0 - c/12)}_{} 
  \boundlk = \left\{ \begin{array}{cll}
  0 & {\rm for} & {\rm J}\,{\star}\,\lambda\,{\ne}\,\lambda \,, \\[.2em]
  \!\!\chilc(2\tau{,}0{,}0)& {\rm for} & {\rm J}\,{\star}\,\lambda\eq\lambda \,.
  \end{array} \right. \eear  \ee 
Here $\bar\varpi_{\rm J}$ is the horizontal part of the fundamental weight of
the relevant affine \lie\ that characterizes the simple
current J and $\chilc$ is a so-called twining character 
\cite{fusS3}\exT{, a generalized character-valued index}.
Similar formul\ae\ hold when one twists in addition by an inner \auto.

\sct{D-branes}
We now focus our attention on \bc s which are relevant to strings and
D-branes. To this end we consider boundaries that respect the full chiral
\alg. The natural structure underlying such \bc s turns out to be
the one of {\em \auto s $\om$ of the fusion rules\/} that 
preserve conformal weights \cite{fuSc6}. The origin of these \auto s is
the freedom that is present in relating the two labels $\lambda$ and
$\Bar\lambda$ of a bulk field $\pho\lambda$, and thus is quite similar to the
origin of the appearence of fusion rule \auto s in the
classification of consistent torus partition functions. But in
distinction to the case of closed \cft, the factorization constraints do not
require that this freedom is fixed in one and the same manner on all surfaces.
Specifically, given a definite torus partition function, which (by taking
\exT{the chiral \alg\ }\chir\ sufficiently large) can be assumed to correspond 
to some fusion rule \auto\ $\pi$, the pairing of $\lambda$ and $\Bar\lambda$ is
as prescribed by $\pi$ on all closed orientable surfaces, but on the disk any
other allowed fusion rule \auto\ $\om$ can appear as well. When $\om\eq\pi$ 
one is dealing with an analogue of {\em Neumann\/} \bc s for free bosons, 
while the counterpart of {\em Dirichlet\/} \bc s of free bosons is given by
$\om\eq\pi\,{\circ}\,\om_{\rm C}$, where $\om_{\rm C}{:}\;\lambda\,{\mapsto}\,
\lambda^{\sss\!+}_{\phantom I}$ denotes charge conjugation.

Note that the choice of $\om$ not only influences the values of the constants
$\Alpha^{AA}_\vac$ and $\rb A\lambda\vac$ in \exT{the relation }\erf{bA}, but 
also the explicit
form of the 1-point block $\boundlk$, which therefore should more precisely
be denoted by $\boundlk_\om^{}$.  Adopting the terminology from the free boson
case, one should refer to the \exT{co-in\-va\-ri\-ants }$\boundlk_\om^{}$ as 
{\em D-brane states\/}, or better as {\em D-brane blocks\/}. In the specific 
case of the theory of $d$ uncompactified free bosons $X^i$ with diagonal
torus \parfu\ and $\om\eq{\rm diag}((+1)^{p+1}{,}(-1)^{d-p-1})\iN{\rm O}(d)$
(acting on the $X^i)$, $\boundok_\om^{}$ is indeed nothing but the usual 
Dirichlet $p$-brane with vanishing field strength on the $p{+}1$-dimensional 
world volume.
The \auto s $\om$ form a group (which in some cases is a Lie group, e.g.\ 
${\rm O}(d)$ for $d$ free bosons). In a space-time interpretation, the choice of
a connected component of that group looks like a topological information; thus
\exT{the \auto\ }$\om$ encodes global topological features of the D-brane.

The choice of a fusion rule \auto\ $\om$ does not refer to a boundary 
of $C$ at all. Therefore this freedom is already present in the absence 
of boundaries, e.g.\ for $C\eq{\dl P}\reals^2$. In contrast, 
as soon as boundaries are present there is an additional freedom, namely 
the \exT{(in general non-unique) }choice of a consistent collection
of reflection coefficients $\rc A\lambda\vac$. Thus a \bc\ $A$
should be regarded as a {\em pair\/} $A\,{\equiv}\,(\om{,}a)$, where $\om$ is
a fusion rule \auto\ respecting conformal weights, while the label $a$ is
tied to the existence of the boundary. In a space-time interpretation, 
$a$ characterizes local properties of the D-brane, such as its position or a 
field strength on it \cite{fuSc6}. In \cite{fuSc6}, $\om$ is called the
{\em\ttype\/} of the \bc, while $a$ is referred to as the {\em\ctype\/}
because in string theory one must attach a distinct Chan\hy Paton 
multiplicity $N_a$ to each allowed \exT{value of }$a$. (The numbers $N_a$ are to
be determined by string theoretic arguments, e.g.\ tadpole 
cancellation.) Note that the summation in \erf{pp} is over all possible
\ctype s $a$ such that $A\eq(\om{,}a)$ with fixed \ttype\ $\om$.

So far we did not say much about the possible values of \exT{the label }$a$.
According to \cite{card9} in the Neumann case $\om\eq\pi\eq\om_{\rm C}$
the allowed index set is equal to the set $\{\lambda\}$ and the associated 
reflection coefficients $\rc A\lambda\vac$ furnish \exT{\onedim\ }\rep s of the
fusion \exT{rule }\alg. In \cite{fuSc6} evidence was collected for the fact
that (for all rational theories, and similarly for certain non-rational ones), 
for fixed \exT{\ttype\ }$\om$ the number of labels $a$ equals the dimension of
some commutative associative \alg\ $\liefont C_\om$ that generalizes the 
fusion \exT{rule }\alg, and that the \exT{reflection coefficients }$\rc A
\lambda\vac$ furnish \onedim\ $\liefont C_\om$-\rep s.
The structure constants of $\liefont C_\om$ are expected to satisfy some
analogue of the Verlinde formula, related to structures similar to those 
uncovered in \cite{fusS3}.
One \exT{particular }class of examples for such classifying algebras had
already been obtained before in \cite{prss3} (for \wzwm s) and \cite{fuSc5}
(for arbitrary \cfts); several other examples are listed in \cite{fuSc6}.
\vfill

\def\wb{\,\linebreak[0]} \def\wB {$\,$\wb}
\def\Bi{\bibitem }
\newcommand\J[5]   {{\sl #5\/}, #1 #2 (#3) #4 }
\newcommand\Prep[2]  {{\sl #2}, preprint {#1}}
\def\nupb  {Nucl.\wb Phys.\ B}
\def\phlb  {Phys.\wb Lett.\ B}
\def\comp  {Comm.\wb Math.\wb Phys.}

\end{document}